# Neuromorphic Time-Dependent Pattern Classification with Organic Electrochemical Transistor Arrays.

**Sébastien Pecqueur[1]\*, Maurizio Mastropasqua Talamo[2], David Guérin[1], Philippe Blanchard[2], Jean Roncali[2], Dominique Vuillaume[1] and Fabien Alibart[1]\***

[1] Institute of Electronics Microelectronics and Nanotechnology (IEMN), CNRS, University of Lille, Av. Poincaré, F-59652 Cedex, Villeneuve d'Ascq, France
[2] MOLTECH-Anjou, CNRS, University of Angers, 2 Bd. Lavoisier, F-49045 Angers, France
\* E-mail: sebastien.pecqueur@*iemn.univ-lille1*.fr; fabien.alibart@*iemn.univ-lille1*.fr

**Based on bottom-up assembly of highly variable neural cells units, the nervous system can reach unequalled level of performances with respect to standard materials and devices used in microelectronic. Reproducing these basic concepts in hardware could potentially revolutionize materials and device engineering which are used for information processing. Here, we present an innovative approach that relies on both iono-electronic materials and intrinsic device physics to show pattern classification out of a 12-unit bio-sensing array. We use the reservoir computing and learning concept to demonstrate relevant computing based on the ionic dynamics in 400-nm channel-length organic electrochemical transistor (OECT). We show that this approach copes efficiently with the high level of variability obtained by bottom-up fabrication using a new electropolymerizable polymer, which enables iono-electronic device functionality and material stability in the electrolyte. We investigate the effect of the array size and variability on the performances for a real-time classification task paving the way to new embedded sensing and processing approaches.**

## 1. Introduction

In biological systems, dynamical and complex information is processed efficiently by highly redundant and parallel network of cells while standard computing systems are quickly reaching their limitations for equivalent information processing tasks. For instance, at the opposite to top-down circuits with highly uniform devices used for general purpose computers, bottom-up assembly of neural cells with high level of variability, can process auditory, visual or olfactive stimuli and generate complex actions very efficiently. Material implementation of such bio-inspired principles for sensing and computing has been a stimulating direction that has reached significant milestones with the development of neuromorphic sensors (retina, cochlea…) and circuits.[1] While initially relying on standard silicon-based devices (i.e. CMOS), emerging materials and devices are opening new avenues for neuromorphic engineering by offering new basic mechanisms for emulating biology and new devices and circuits concepts to build computing systems. Notably, neuromorphic systems with non-volatile memories (and resistive memory in particular) have been the focus of strong research efforts.[2]

Here, we capitalize on organic electro chemical transistors (OECTs) that have been recently proposed ubiquitously as basic building blocks in neuromorphic computing applications[3] (i.e. memory devices, for instance) and as bio-sensors thanks to their intrinsic sensitivity to ions. For example, we recently reported how OECT devices can be used for discriminating ions based on transient responses of the ionic transistor subject to pulse stimulation.[4] While this paper does not present sensor properties assessment, we build on this intrinsic feature of OECT to demonstrate neuromorphic computing with potential applications in bioelectronics (see below). Based on an array of organic electrochemical transistors, this work shows how sensing and processing can be realized at the interface with an analyte by taking advantage of OECTs intrinsic physics and on neuromorphic concepts. In particular, we show how highly variable material engineering routes that are not adapted to standard information processing technologies (i.e. relying on top-down fabrication of near-ideal components and circuits) can be turned into an advantage when bio-inspired concepts are used to engineer computing system.

We propose an adaptation of the recent proposition of reservoir computing (RC),[5,6] to demonstrate that both sensing and computing can be obtained from the intrinsic properties of a transistor array, limiting the separation between these two elementary levels (i.e. sensing and computing). On the one hand, from the neuromorphic computing side, RC concept has been developed for dynamical signal processing (e.g. speech recognition),[7] and use the idea of learning from a simple read-out layer (i.e. a feed-forward perceptron) the dynamics associated to the projection of a given stimuli into a complex and random network of non-linear elements (i.e. neurons or nodes). On the other hand, from the sensing perspective, monitoring and analyzing biological activity in medium such as neural cells assembly or bloods composition, for instance, consist in processing dynamical signals and would strongly benefit from the RC approach to classify such dynamical patterns from complex and poorly define biological medium. Thus developing RC strategies to process information out of an ion-sensitive transistors network could open new perspectives for biological sensors.

The key elements of RC are (i) non-linearity of the reservoir's nodes (non-linear conversion from input signal to output signal) and (ii) a fading memory effect keeping the history of the stimuli active in the network on a given duration. Material implementations of RC have been proposed recently with optical or magnetic oscillators.[8-12] For both, time multiplexing was used in order to emulate spatial nodes in the network from one single non-linear element and memory effect was associated whether to a feedback loop connection or to the transient dynamics of the non-linear element. Here, we propose the implementation of a spatial reservoir composed of an array of OECTs that present a non-linear response to the stimulus propagating in an analyte (an input voltage applied to the analyte is converted into a resistance state of the OECT. The memory effect is associated to the transient dynamics of ions penetrating into the OECT with a given relaxation time implementing the fading memory. We show in this paper that this spatial reservoir take advantage of (i) the variability in the OECTs array inherent to the bottom up fabrication of the OECTs using a newly synthesized electropolymerizable polymer, (ii) of the transient dynamics of the devices for an implicit representation of time and (iii) of the number of redundant OECTs to discriminate simple dynamical patterns.

## 2. Results and Discussion

### 2.1. Transient dynamics of OECTs as implicit time representation

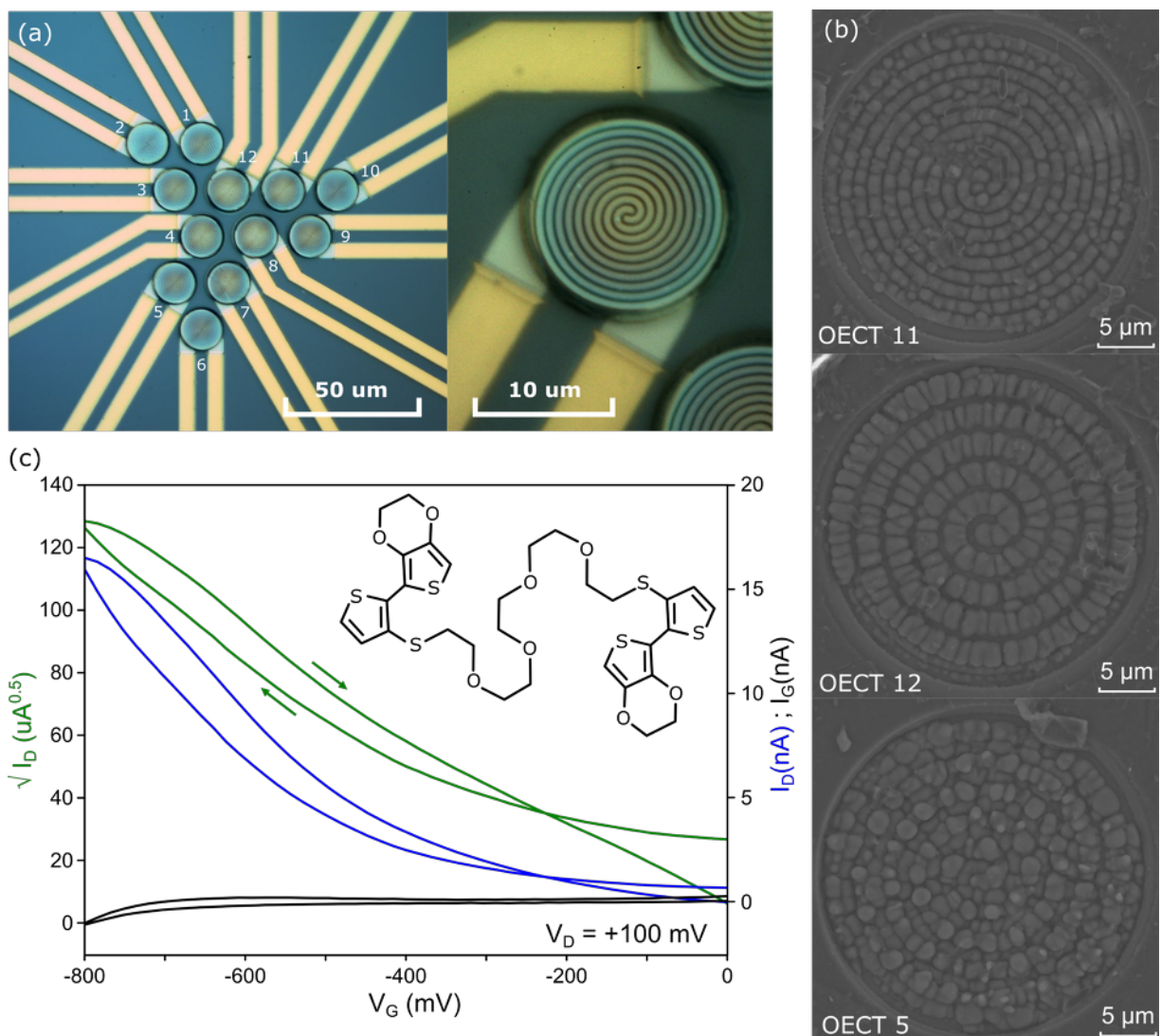

***Figure 1.*** *a, Optical micrographs of the OECTs array. OECT devices are three terminal devices with spiral-shape S and D electrodes leading to a large channel width over length ratio (W/L=1100) over a confined area (615 μm²). The gate electrode (not represented) is realized with macroscopic metal wire contacting the electrolyte. b, After electrode patterning, the organic material, Scanning Electron Microscope images of three different devices. High variability in the electro-polymerization is apparent in the polymer structure. c, Transfer characteristics of one of the OECT in a 0.1 M KCl(aq) electrolyte, displaying the p-type accumulation-mode field-effect (inset: TEDOT monomer used for the semiconductor electrodeposition). Blue line: drain current, green line: square-root of drain current, black line: gate current, showing a negligible capacitive contribution in the overall drain current.*

The OECT array (**Figure 1**a, and Methods) consists in organic electrochemical

transistors composed of electro-polymerized, glycol crosslinked, 2-(2-thienyl)-(3,4-

ethylenedioxythiophene) (TEDOT) molecules (Figure 1 and Supplementary Figure S1). Alternative materials based on glycol-side-chain polythiophene have demonstrated high performances while operating in accumulation mode.[13,14] We used the monomer TEDOT (Figure 1) to conceive, after electro-polymerization, a new polythiophene functionalized with ethylene glycol chains, patterned locally on each OECT. After the polymer electrodeposition (see Methods), all devices showed gate modulation of source-drain current (Figure 1c), despite the large variability of the material morphology (Figure 1b). The polymer formed by nucleating first on the source and drain electrodes and then growing and covering the channel. Such bottom-up fabrication did not resulted in smooth thin-films, but rough patches of polymer grains with heights of about 2 μm, filling the Parylene C cavities. Such textured surfaces form large interfaces with the electrolyte and the relation between the aspects of the patch with the ion dependent electrical characteristics will be discussed further. The basic mechanisms of OECT is based on the redox doping/dedoping of the organic material.[15,16] A negative gate voltage ($V_G$) applies to the 0.1 M KCl$_{(aq)}$ (the analyte) forces negative ions to penetrate into the organic material, increasing the electronic conductance of the organic layer (source grounded and drain potential constant $V_D$=+100 mV). When $V_G$ is turned off, ions diffuse back to the electrolyte, out of the organic material that recovers its high resistance state. Non-linear relationship between $V_G$ and device's resistance is evident from Figure 1b. In addition, slow dynamics of ions through the electrolyte/organic interface is apparent in the hysteresis loop when sweeping $V_G$ at 0.1 V/s. **Figure 2**a presents the transient response of an OECT to a sequence of pulses with increasing amplitude. The transient behavior of the OECT is used to implement short-term memory effect (Figure 2c),[17,18] as observed in biological synapses (Short-Term Plasticty).[19] When the OECT is stressed with a train of pulses of constant amplitude, short-term facilitation (increase of the average output current with number of pulse) is implemented.[3,17]

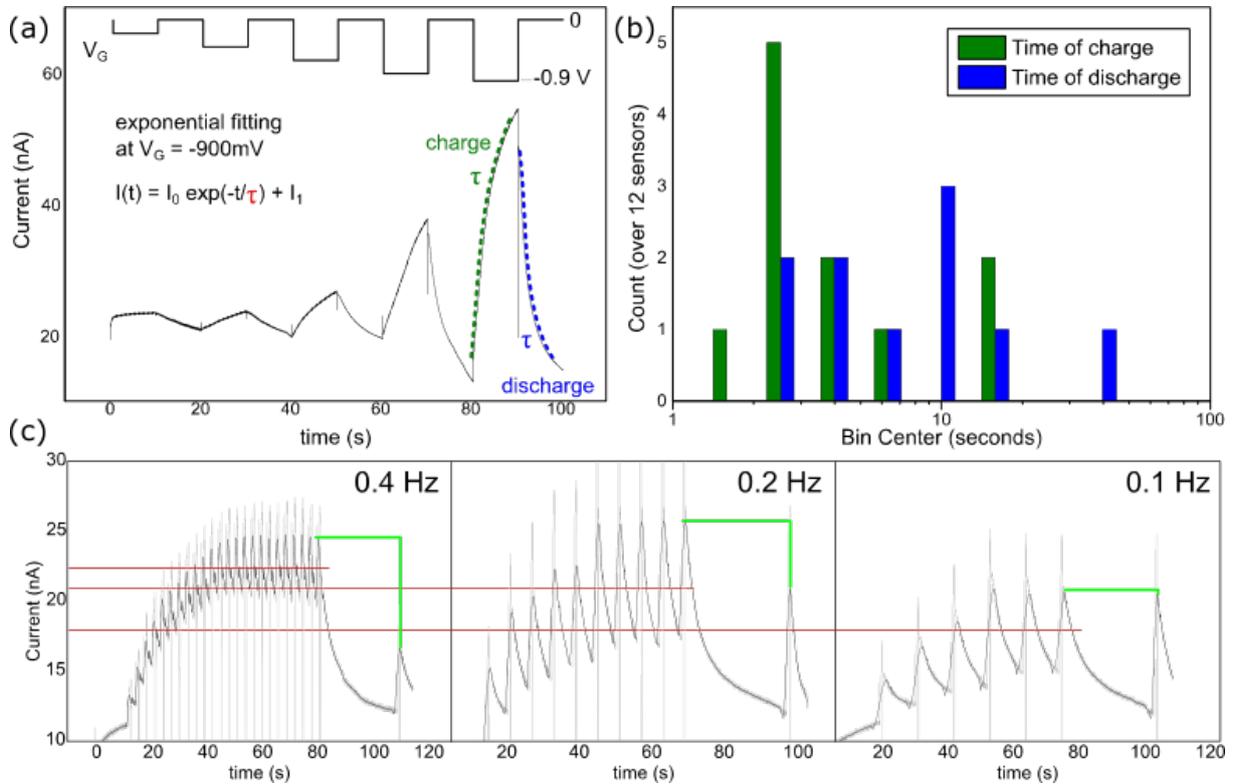

***Figure 2.*** *a, Typical response of the OECT to a gate voltage with different amplitudes and constant $V_D$=+100mV. Green (blue) dashed line shows the fitting of the doping (dedoping), transient resulting from ions injected (removed) to (from) the organic semiconductor for a step voltage of 0 to -900 mV (-900 mV to 0 V). b, Characteristic charge and discharge time constants for the full array of 12 OECTs obtained from fitting the transient current in (a) by exponential functions (dashed lines in (a)). c, Constant gate voltage stimulation (-900 mV, 200 ms width) with variable frequencies showing short-term facilitation. Red dashed lines correspond to the average conductance reach in steady-state showing frequency-dependent potentiation. Short-term memory effect is evidenced by the single pulse applied 25 seconds after the potentiation. Green lines are guide to the eyes to evidence the stronger potentiation/relaxation ratio after higher frequency stimulation.*

This short-term memory effect is due to the accumulation of ions from pulse to pulse and on the unbalanced relaxation between each stimulation resulting in higher steady-state mean conductance at high frequency (larger amount of ions are accumulated) and lower steady-state conductance at low frequency (lower amount of ions are accumulated). In each case in Figure 2c, a single control pulse is applied after 25 s of relaxation to evidence the short-term memory effect. The absolute value of the current from trial-to-trial presents some variability in current level (evidenced in the very first pulse when the OECT was previously at rest, for instance) that was mostly due to concentration control of the electrolyte micro-drop in an open environment and to the strong effect of the concentration on the quantitative response of the OECT. Nevertheless, the qualitative response showing higher potentiation at higher frequency with respect to the control pulse (green lines in Figure 2c) remained consistent. The transient behavior was analyzed by fitting the charge/discharge in the transient drain current characteristics of Figure 2a with a single exponential function. Characteristic times with a large dispersion are obtained (Figure 2b), inherent to the bottom-up fabrication process of the organic material. Characteristic transient times in an OECT depend on the electrical resistance of the polymer and the resistance and capacitance values of the electrolyte,[16] affecting the device behavior under pulse modulation.[4] Since the polymer and electrolyte resistances as well as the device-to-electrolyte capacitance are intrinsically function of the thickness of the polymer materials and the areas of their interfaces, the variability of polymer morphology (as shown in Figure 1b) is the main source of variability of the measured device time constants. This will be one central element that we exploit in the following for the implementation of RC. Characteristic time constants for charging are on average shorter (from 1.08 to 13.7 s) than discharging (from 2.03 to 43.7 s). The level of memory of each individual device can be define along two metrics: (i) $\tau_{mean}$, the average value of $\tau_{discharge}$ and $\tau_{charge}$ and (ii) the $\tau_{discharge}/\tau_{charge}$ ratio. When $\tau_{discharge}/\tau_{charge} \approx 1$ (equivalent to a capacitor), the device is a purely short-term memory. When $\tau_{discharge} / \tau_{charge}$ tends to higher values, the memory moves from short-term to long-term memory (note that non-volatile memory tends to maximize this ratio with $\tau_{discharge} > 10$ years and $\tau_{charge} < 1$ ns). Figure S5 and S6 represents the $\tau_{discharge}/\tau_{charge}$ ratio and $\tau_{mean}$ showing that OECTs devices are in the short term regime of memory. In the following, this short term memory effect will be used to define the global memory of our system. The collection of $\tau_{mean}$ available in our system due to variability will be used to reconstruct a memory time window. This characteristic memory time window directly determines the typical duration of dynamical patterns that can be processed by the reservoir of OECTs (i.e. a device can keep memory of its previous history on a time window of up to tenths of seconds). In its initial version, RC used recurrent connections into the reservoir to implement fading memory effect. Feedbacks (i.e. delays) into the reservoir ensure that signals are kept for a given time active in the network. Also, strength of this recurrent connections was used to set the reservoir in an optimal state in terms of sensitivity to input signal (i.e. edge of chaotic regime).[20] In our case, the reservoir consists in a purely feed-forward network (i.e. no recurrent connections) and fading memory effect is implemented with the transient responses of OECTs (more precisely by the collection of time constant from each individual OECT). The optimal $V_G$ range of operation of the OECTs is defined based on the device $I_D(V_G)$ characteristics. Too large voltage biases (>1.0 V) might lead to irreversible material damage by water electrolysis,[21] hindering the stability of the electrochemical system. Too small voltages result in too weak modulation of the conductance. As a trade of, we use voltage pulses $V_G$ of -900 mV. Note that normally-OFF OECT (i.e. intrinsic semiconductor) require a substantially larger gate voltage in order to achieve a reasonable doping of the channel at the opposite to normally-ON OECT (i.e. doped semiconductor such as PEDOT:PSS) that can be operated at ultra-low voltages.[3]

## 2.2. Reservoir computing: dynamical signal processing with network of OECTs

Various Artificial Neural Networks (ANNs) approaches have been proposed so far for time-dependent pattern classification (speech, for example). Recurrent networks or Time Delay Neural Network are of particular interest for this task since they offer the possibility to encode the time signature of such signals explicitly.[22] More recently, time-dependent signals processing has been revitalized with the RC concept. RC is based on the basic idea of projecting the input signal on the nodes of a large dimensional space in order to separate simple features from the input signal. These simple features are then used to classify patterns at the read-out layer (i.e. a simple perceptron trained with standard learning technics). In time-multiplexed RC approaches used for speech recognition,[8,11] the dimensionality of the reservoir is ensured by the virtual neurons (i.e. virtual nodes) that hold the signature of the signal at different time intervals. Reconstruction on the read-out layer of the time-dependent signals out of these virtual nodes is then used to classify patterns (i.e. speech signals). This approach corresponds to an explicit representation of time where the first neuron is associated to the first time interval, second neuron to the second time interval… and so on.[22] Here, we use an implicit representation of time through the transient dynamics of each OECT in the network (**Figure 3**a). Due to the variability in their transient responses (Figure 2), each OECT will keep the temporal signature of the signal on a different memory window. Each OECT is then used to collect different features from the reservoir and to perform classification at the read-out layer.

To test this concept, we designed low complexity signals consisting in square waves of constant amplitude -900 mV and 1 s duration applied to the global gate with variable frequencies. The two signals used to demonstrate time-dependent signal classification are built with square-type and triangle-type pulse-frequency modulation between 0.3 and 0.8 Hz.[23] If classification of these two signals is trivial when one have access to the full recording over a complete period of the signals, discrimination of the two signals on a restricted time interval (typically no more than two successive pulses) becomes impossible without some memorization of the past events. Here we show that the RC concept can be used to classify in real time these signals based on the intrinsic memory of each OECT and on pre-requisite learning. Figures 3b-c present the typical response of an OECT to the two signals, respectively. Light grey lines correspond to the as recorded signal. The blue and green lines correspond to the average current response in each time step.

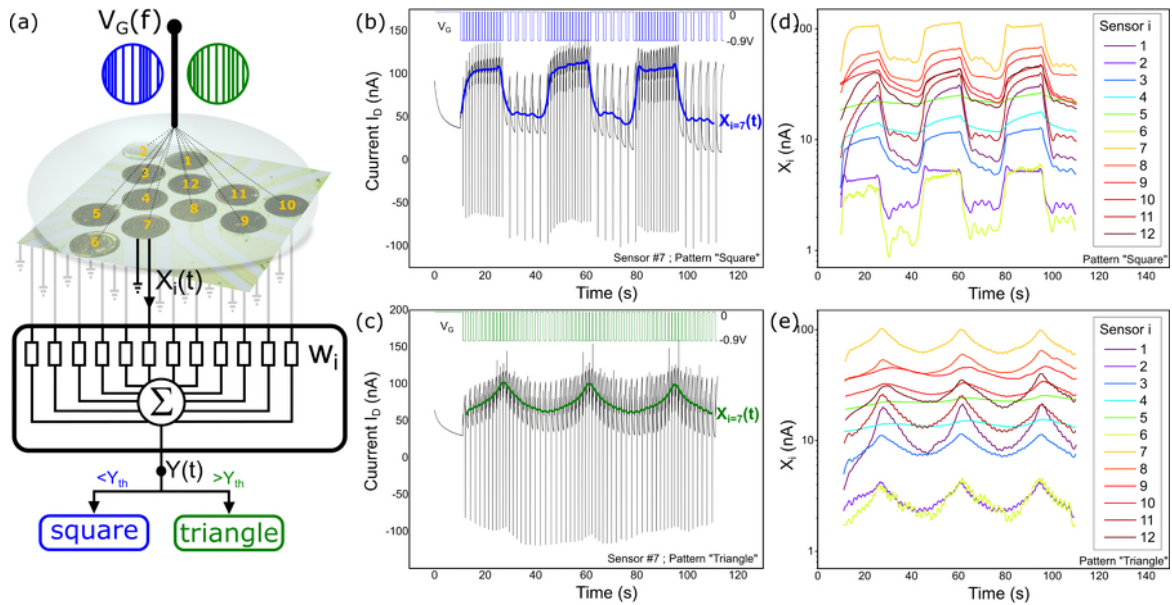

**Figure 3.** *a, Schematic of the experiment and analysis. The network of 12 OECTs is used to sense the response of the electrolyte to the global gate signal. The perceptron is implemented in software and realize signal weighting, summation and activation function. b and c, Typical response of an OECT to the two different patterns (square and triangle, respectively) used for classification. Black/grey lines are the raw data measurement and blue/green line are smoothed signals corresponding to the average current value sense by the device. As in Fig. 2c, highest (lowest) frequencies tends to accumulate more (less) ions in the organic material and increase (decrease) the mean conductance levels of the OECT via doping effect. d and e, Response of the 12 OECTs to the two input patterns (square type and triangle type, respectively). We observe variability from device to device on the modulation amplitude and the shape of the mean current. Average currents of OECT #6 and #7 differ of about two orders of magnitude, and amplitude of the current modulations between OECT #1 and #4 differ of about one order of magnitude.*

Figures 3d-e present the responses of the 12 OECTs in the array to the same signal applied at the common gate. We observe variability from device to device on the modulation amplitude and the shape of the mean current. The observed variability on the mean current is not only due to the $\tau_{discharge}$ and $\tau_{charge}$ device time constants variability, but also to their steady state current and its current modulation. Both of them correlating to the hole conductivity of the polymer between source and drain electrodes, these two features being highly influenced by the polymer morphology: Comparing the patches morphology (Figure 1b) to the device performances (Figure 3d-e), small-island morphological features of the OECT 5 seems to promote poor current modulation (ion-gating poorly influencing the conductive region of the channel), and a 1-electrode nucleation mode (OECT 12) seems to promote the average drain current by a factor of three compared to a 2-electrode nucleation mode (OECT 11), forming no grain boundary in the middle of the channel. This large variability represents a severe limitation of bottom-up fabrication technics that RC can leverage efficiently. Implementation of the reservoir target classification of the two patterns (square-type and triangle-type) with a simple read-out equivalent to a simple perceptron implemented here in software (i.e. one neuron with m=12 weighted input). Output current from each OECT has been sampled over time and will be used to define the state of the reservoir at each time step, for a given location into the reservoir. The collection of outputs from the array of OECTs correspond to the state at time t of the reservoir in response to a given stimuli. This output values are then used as input to a simple perceptron in charge of classification through learning (i.e. the perceptron function is limited to signal weighting, summation and activation function). We use a 1 ms sampling rate of the signals. For each time step, we associate a given vector $\{X_i\}(t)$ of dimension (1x12). A total of 11750 vectors are recorded from each pattern composed of three

repetitions of one period of square (triangle, respectively) elementary pattern. Each vector is then fed to a simple perceptron with m=12 weighted inputs. The total output Y(t) from the perceptron before activation function at time t is then

$$Y(t) = \sum_{i=1}^{m=12} w_i \cdot X_i(t) \qquad (1)$$

with $w_i$ the synaptic weight of the i[th] input line, $X_i(t)$ the current value at time t of OECT #i. Vectors $\{X_i\}$ belongings to the triangle-type pattern are associated to class "1" (i.e. output neuron is activated) and square-type pattern to class "0" (i.e. output neuron is not activated). The activation function rule of the output neuron is then

Output = "1" if Y(t) > $Y_{th}$
Output = "0" if Y(t) < $Y_{th}$

with $Y_{th}$ the threshold activation value (0.5 in our example) chosen based on the distribution of the different output values. We define the training vectors (or training examples) by choosing randomly n vectors from each pattern. Training protocol is realized with pseudo-inverse learning (i.e. Moore Penrose operator)[24] to determine the value of the 12 synaptic weights (Supplementary Figure S2). Testing is realized on the full set of vectors (i.e. 11750) from the two classes. Classification performances are then calculated as the percentage of errors averaged on 20 iterations with 2n training vectors randomly chosen among the entire set. **Figures 4**a-b show the error rate as a function of training vectors used for learning the synaptic weights. As in standard learning technics, more examples lead to better performances. With more than n=300 vectors (only 2.6% of the total vectors set), the system reaches classification of the signal with an error rate at the order of 0.001%.

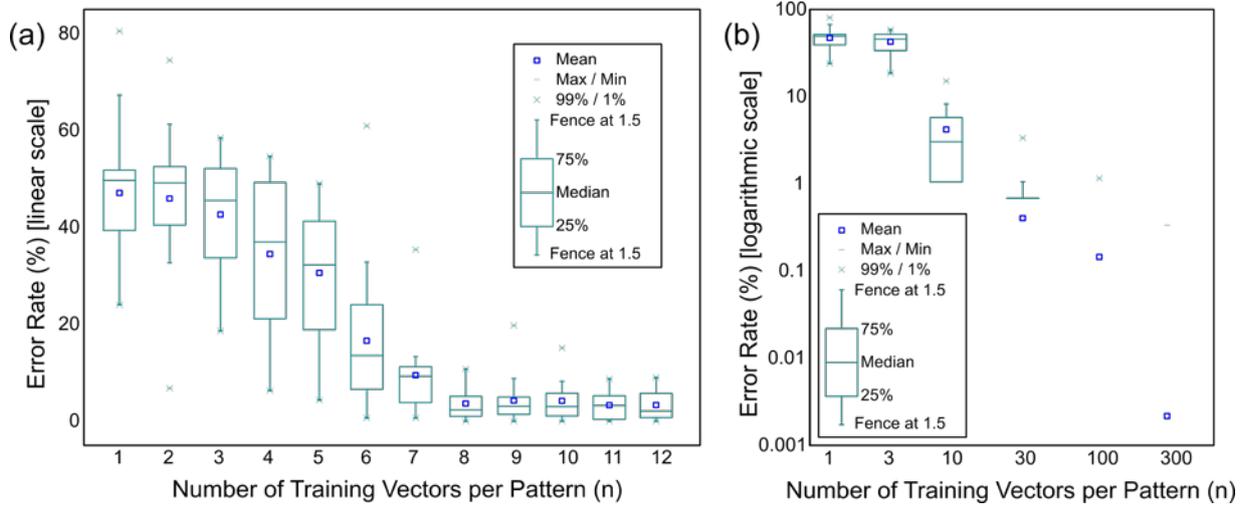

**Figure 4.** *Error rate (i.e. the number of vectors given a mistaken recognition over the total number of tested vectors) for classification of the two patterns as a function of number of training vectors n in each class. Testing is performed on the entire vector set (11750 vectors). a, linear scale. b, logarithmic scale.*

The present classification task is a toy problem and any quantitative discussion on the error rate is pointless. The important aspect here is to show that the system after learning can classify in real time input signals since the only operation to be realized out of the OECT array is current weighting, summation and thresholding (excluding the learning stage which require a significant initial computing power and time). Note that we used the average value of current that can be easily implemented in hardware with a simple integration circuitry.

These results are based on two important aspects. (i) The number of OECTs used for pattern classification that we have associated to the number of features collected from the reservoir. (ii) The variability from device to device that affects how much each feature is different from the other. For instance, two identical OECTs will provide the same feature while very different OECTs will provide very different features (i.e. different transient time between two devices will provide different memory window and consequently, different type of features). To demonstrate that RC takes directly advantage of both number of features and variability in the OECT array, we realize the same classification task with only "partial" arrays of m OECTs, with $1 \leq m \leq 12$.

### 2.3. Influence of the number of OECT in the reservoir

Performance of classification should be directly linked to the number of features (associated to each OECT) used to classify the patterns. To test this hypothesis, we evaluate the performances of the reservoir, degrading it on-purpose by removing sequentially OECTs one by one. **Figures 5**a-b present the error rate as a function of the number of training vectors n when one OECT is removed from one batch to the other.

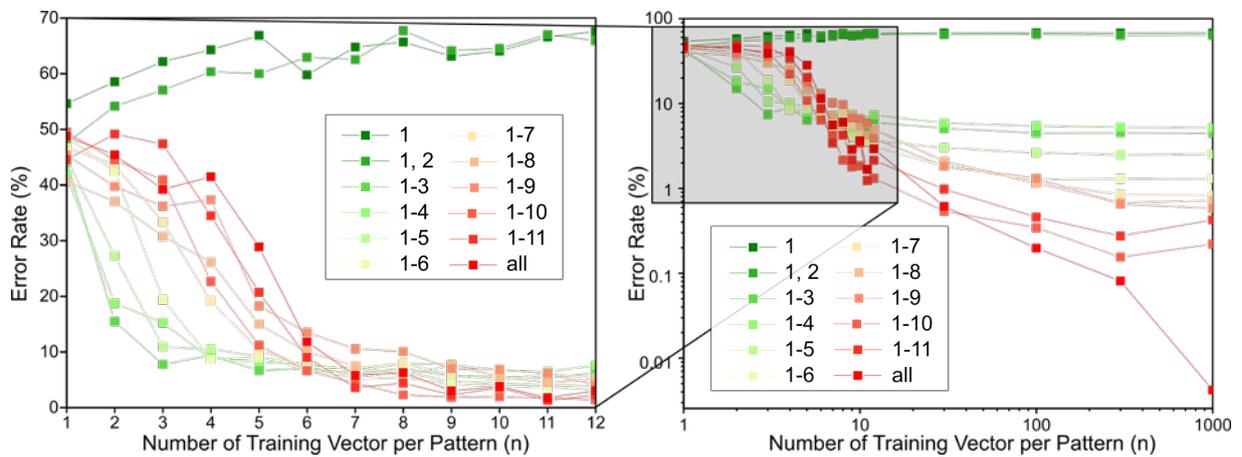

***Figure 5.*** *Error rate of the classification task when the OECTs' array is progressively degraded by removing sequentially OECTs one-by-one.*

A minimum of three OECTs is required to classify patterns. When more than 300 training vectors are used, a clear relationship between number of OECTs and performances is extracted with the later improving exponentially with the number of OECT used to classify the pattern. Performances of classification are directly linked to the number of OECT used to compute the pattern out of the reservoir, or equivalently, to the number of features used to describe the patterns. Referring to the general idea of RC, increasing the number of OECT in the network corresponds to increasing the dimensionality of the space used to project the input signal. The expected effect is consequently to ease the readout layer (i.e. the perceptron) to classify the patterns. In this sense, more OECTs are better for classifying time-dependent signals. Classification of more complex patterns (i.e. patterns with higher level of similarity, for example) should require more OECTs in the array. We notice that small arrays (m<6) require less training vectors to reach error rate of about 10%. In agreement with theoretical prediction from ref. [25], this effect implies that for a relatively small number of training vectors, it exists an optimal number of features to reach the best accuracy. In other words, it shows that the training shall be adapted to the size of the OECT array, which should be adapted to the task complexity.

### 2.4. Influence of the variability in the reservoir

Another important issue is to know what type of features could lead to better performances. **Figure 6**a shows the error rate for the particular case of 6 OECTs out of 12 when all combinations (i.e. 924) are tested. Figure 6a clearly shows that some set of OECTs perform better than other, but it was not possible to extract a possible empirical rule correlating the error rate to the level of variability in transient dynamics and/or mean current level (see Supplementary Figure S4 for details). Nevertheless, important insights are obtained by considering the average performances of the array as a function of number, m, of OECTs. Figures 6b-c show the error rate for different sizes of array (m) obtained with n=3 (small training set) and n=300 (large training set) training vectors. Each value is obtained by extracting the mean error rate from 10 randomly chosen sets of OECTs (the same sets for both n=3 and 300). Figure 6b confirms theoretical prediction[25] that there exists an optimal error rate performance for a finite number of training vectors n. This effect attenuates when the number of training vectors increase and disappears for n>6 (Figure 6c) where performance improves monotonically with the number of features when n=300 (Supplementary Figure S3). We can speculate at this stage that: (i) the propose concept requires variability. One or two OECTs are not enough to perform classification at a better level than chance. Since the output of each OECT is weighted by the perceptron (i.e. corresponding to synaptic weighting, signal summation and activation function), 12 OECTs without variability and providing the same response would be strictly equivalent to a perceptron with a single weight. This system, equivalent to a Boolean logic gate, cannot be used for classification task. (ii) The reservoir concept is rather resilient to the nature of variability presents in each individual OECT since absolute performance as a function of m is larger than mean deviation for each m values (see error bars in Figure 6b). In other words, the number of OECTs seems to play a more critical role on performances than the possible effect of a particular set of OECT (i.e. a particular set of features). The consequence of this point is to consider that increasing the number of OECTs allows coping with uncontrolled variability. More insights about the nature of variability should require higher OECTs' array with more complex input patterns in order to extract relevant trends between variability and performances.

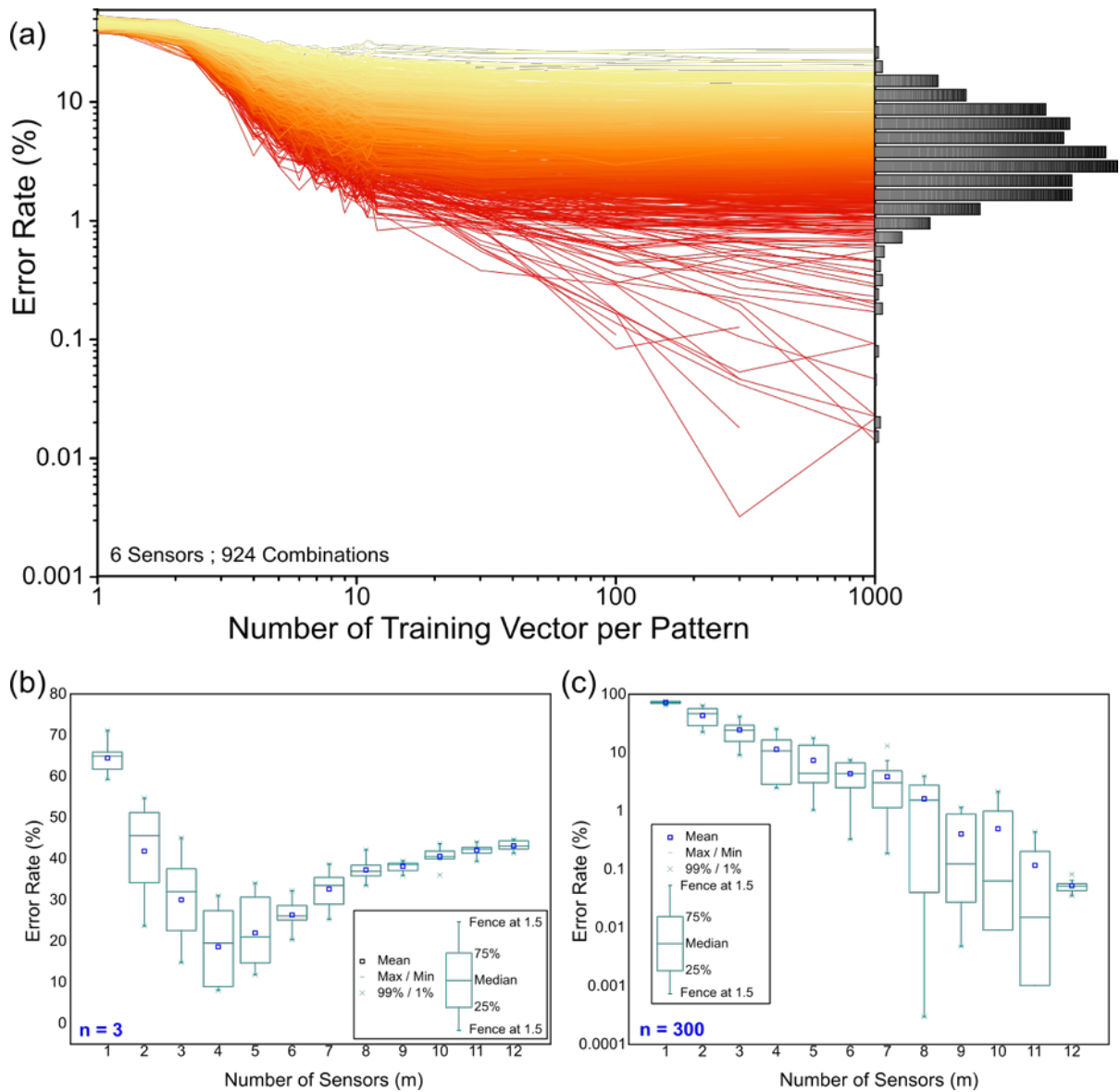

***Figure 6.*** *a, Error rate of the classification task obtained for all combination possible out of the 12 OECTs of the array as a function of number of training vectors. Performances for different array size m with b n=3 (b) and n=300 (c) training vectors. Average performances as a function of m is obtained by calculating the mean value and deviation on 10 randomly chosen set of vectors for each m value (vector set displayed in Supplementary Table S1).*

## 3. Conclusion

We demonstrated in this study that despite the high level of inherent variability in our bottom-up ion-sensing devices, we successfully discriminated dynamic patterns out of our OECT network by using a well-established neuromorphic learning algorithm. This approach could be applied to various practical cases since most of biological processes such as electrical activity in brain cells or evolution of composition in physiological medium belongs to the class of dynamical patterns. We showed that the RC approach can efficiently cope with many-fold variabilities in both transistor characteristic time constants and non-linear current levels to recognize frequency-modulated pulsed signal. Although it was not possible to correlate the dispersion of each device properties with the recognition performance of the OECT array, this work shows that neuromorphic sensing takes advantage of these variabilities, considered as drawbacks for standard sensing. While today's transistor technologies are based on high reproducibility and fast response, we point out the potential to rethink the excellence criteria for sensing in a neuromorphic data analysis context, and open up possible directions for future sensing technologies based on variability-rich materials, grown by bottom up approaches.

## 4. Experimental Section

*Monomer Synthesis:* The monomer TEDOT has been synthesized by deprotection/functionalization of the appropriately protected thiolate groups according to the already published method,[26] and will be described elsewhere. The identity and purity of TEDOT were confirmed by $^1$H and $^{13}$C NMR spectrometry and HR mass spectrometry giving satisfactory results: Yellowish oil. $^1$H NMR (300 MHz, CDCl$_3$, δ/ppm): 7.21 (d, $^3$J = 5.3 Hz, 2H, H$_{thioph}$), 7.03 (d, $^3$J = 5.3 Hz, 2H, H$_{thioph}$), 6.36 (s, 2H, H$_{EDOT}$ $_{Ar}$), 4.37 – 4.21 (m, 8H, HEDOT Alk), 3.63 – 3.51 (m, 16H), 2.98 (t, J = 7.1 Hz, 4H); $^{13}$C NMR (76 MHz, CDCl$_3$, δ/ppm): 141.32, 139.14, 134.73, 132.30, 126.13, 123.82, 110.85,

99.82, 70.78, 70.70, 70.46, 70.29, 65.17, 64.59, 35.74; HRMS (FAB, m/z): calculated for $C_{30}H_{34}O_8S_6$: 714.0578; found: 714.0581.

*Device Fabrication:* The fabrication of the clusters of 12 OECT devices was based on methods already reported in the literature:[27,28] We first patterned on a Si/SiO$_2$ substrate the 70 nm Pt source and drain electrodes, and then the 300 nm Au electrodes (with a 10 nm Ti adhesion layer) by e-beam lithography and lift-off using a PMMA/MAA resist. Pt was chosen as a metal for the source and drain electrodes for its electrochemical stability, while gold was used for its higher conductivity to reduce the resistance of the transmission lines on the substrate. After UV-O$_3$ cleaning, the substrate was functionalized with 3-acryloxypropyltrimethoxysilane in o-xylene (with 1% acetic acid) before processing a 2 μm layer of Parylene C by chemical vapor deposition, as a passivation layer. The apertures over the active area of the devices were realized by lithography, using an AZ40XT resist (from MicroChem) and dry etching with an O$_2$ plasma. The polymeric semiconductor p(TEDOT) was deposited sequentially over each device by electrodeposition, using the source and the drain electrodes as working electrode, a hanging Pt wire as a counter electrode and an Ag/0.01 M AgNO$_3$ reference electrode. The potentiostatic electrodeposition (perform on a Solartron Analytical ModuLab 2100A) was performed at 1 V vs. Ag/Ag$^+$ until 1 μC over each electrode was deposited, using a $10^{-2}$ M solution of TEDOT in acetonitrile in the presence of 0.1 M of tetrabutylammonium hexafluorophosphate as electrolyte. The patches of p(TEDOT) adhere remarkably on the device upon intense rinsing with deionised water prior the electrical tests.

*Electrical Characterization:* The squared voltage inputs were applied via three Agilent B1530A waveform generator units. All measurements were performed using a silver wire as a gate, dipped into a 0.1 M KCl$_{(aq)}$ electrolyte. Current-voltage curves were measured with Agilent B1530A.

## Supporting Information

### Acknowledgements

We acknowledge financial supports from the EU: H2020 FET-OPEN project RECORD-IT (# GA 664786). We thank the French National Nanofabrication Network RENATECH for financial support of the IEMN clean-room. We also express our thanks to Christian Gamrat (CEA, Saclay) for fruitful discussions.

### Conflic of Interest

The authors declare no competing financial interests.

### Author contributions
S.P. and D.G. fabricated the devices. M.M.T., P.B. J.R. designed and synthesized the electropolymerizable polymer. S.P. performed the measurements and analyzed the data. S.P and F.A. designed and performed the RC training. D.V. managed the CNRS side of the EU RECORD-IT project. S.P., F.A. and D.V. wrote the paper. All the authors discussed the results and contributed to correct and amend the paper.

### Keywords

Biomimetics, Conducting Polymers, Organic Electronics

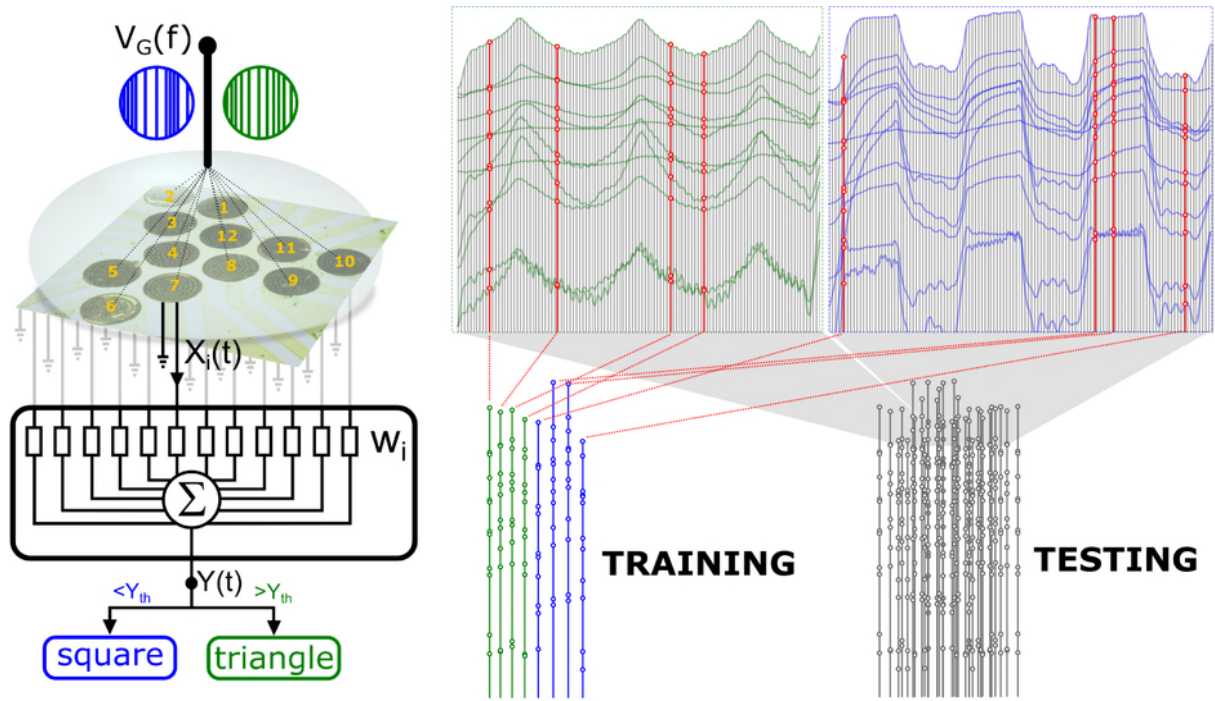

## ToC

Hardware reservoir-computing realized with an array of micro organic electrochemical transistors. Demonstration of the possibility to classify frequency-modulated voltage-pulse patterns and the possibility to scope with the rich morphological variability of the 12 electropolymerized semiconducting devices on which the gate input is projected.

# Neuromorphic Time-Dependent Pattern Classification with Organic Electrochemical Transistor Arrays.

## *Supplementary Information*

**Sébastien Pecqueur[1]\*, Maurizio Mastropasqua Talamo[2], David Guérin[1], Philippe Blanchard[2], Jean Roncali[2], Dominique Vuillaume[1] and Fabien Alibart[1]\***

[1] Institute of Electronics Microelectronics and Nanotechnology (IEMN), CNRS, University of Lille, Av. Poincaré, F-59652 Cedex, Villeneuve d'Ascq, France
[2] MOLTECH-Anjou, CNRS, University of Angers, 2 Bd. Lavoisier, F-49045 Angers, France
\* E-mail: sebastien.pecqueur@*iemn.univ-lille1*.fr; fabien.alibart@*iemn.univ-lille1*.fr

**Device fabrication**

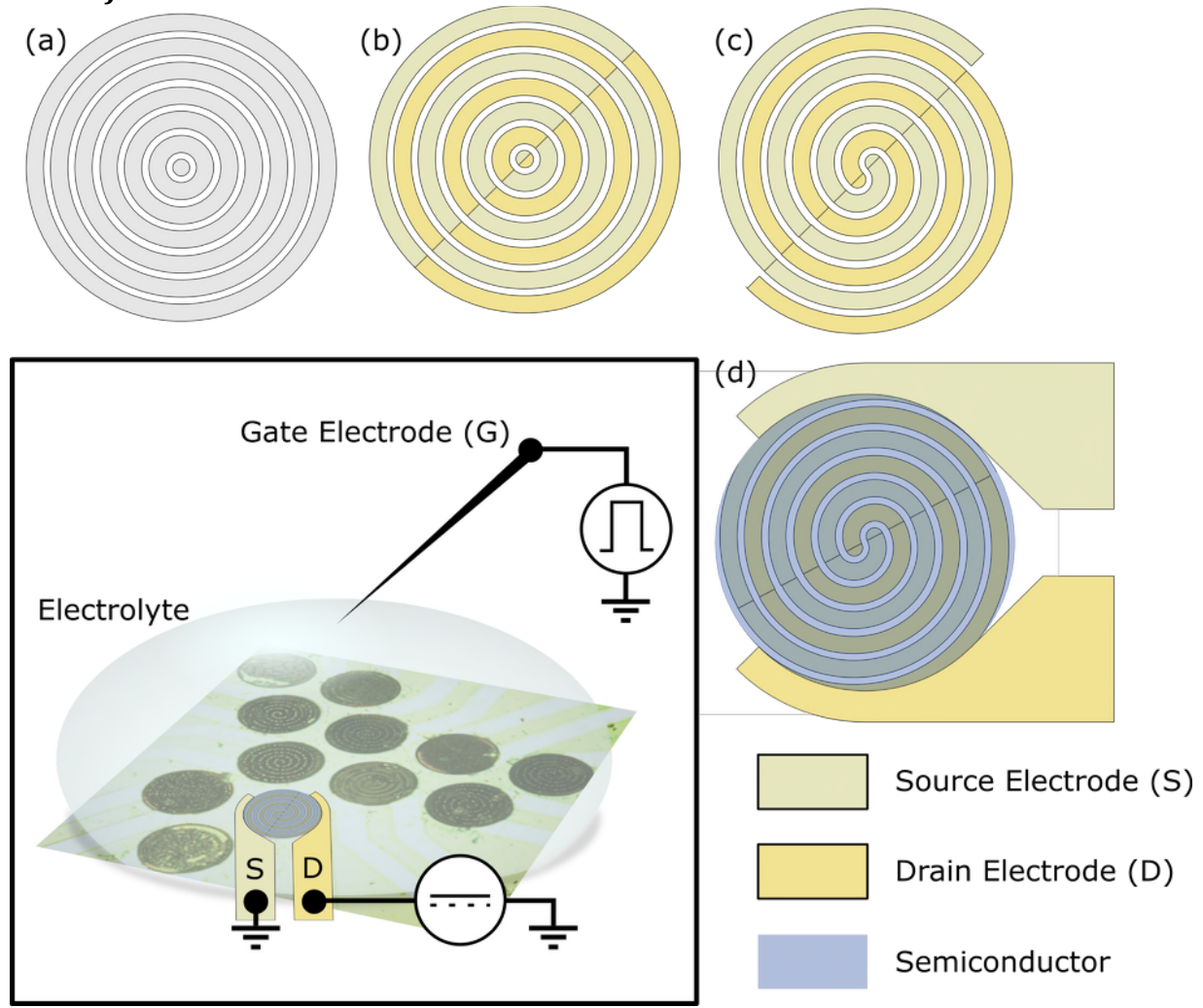

**Figure S1.** *Strategy to prepare the concentric spiral-like source and drain electrode geometry: a, The spiral-like interdigitated electrodes we designed from concentric rings (ring width W = 1 µm; interspace L = 0.4 µm). b, After splitting the concentric rings into two halves and c, translating them along the cut axis by W+L, one obtain this C2 symmetry for which the channel length L is controlled over the whole structure. d, Both spiral electrodes are contacted on each side of the active area to supply the voltage $V_S$ and $V_D$. In the squared box, a schematic of the electrical setup for operating the OECTs.*

### Pseudoinverse Learning (PIL) algorithm

We picked up randomly (using the rand() function of Microsoft® Excel™) 2n vectors of m transistors outputs for training: n from the group of outputs for a "square" signal and n from the group of outputs for a "triangle" one.

We defined the training matrix $X=\{x_i{}^j\}$, i from 1 to m and j from 1 to 2n and the 2n-dimensional vector $Y=\{y_i\}$, such as for each i from 1 to 2n, $y_i=0$ if "triangle" or 1 if "square".

We looked for the weight vector $W=\{w_i\}$ by computing $X^+=(X^t \cdot X)^{-1} \cdot X^t$, pseudoinverse matrix of X, in order to find $W=X^+ \cdot Y$.

We evaluated $y_i=\{x_i\}^t \cdot W$ for every m-dimensional vectors acquired during both "triangle" and "square" stimulation: If $y_i<0.5$, we assumed the algorithm recognized a triangle while if $y_i>0.5$, it recognized a square. We evaluated the error rate by averaging a minimum of 20 iterations of this algorithm for each data point.

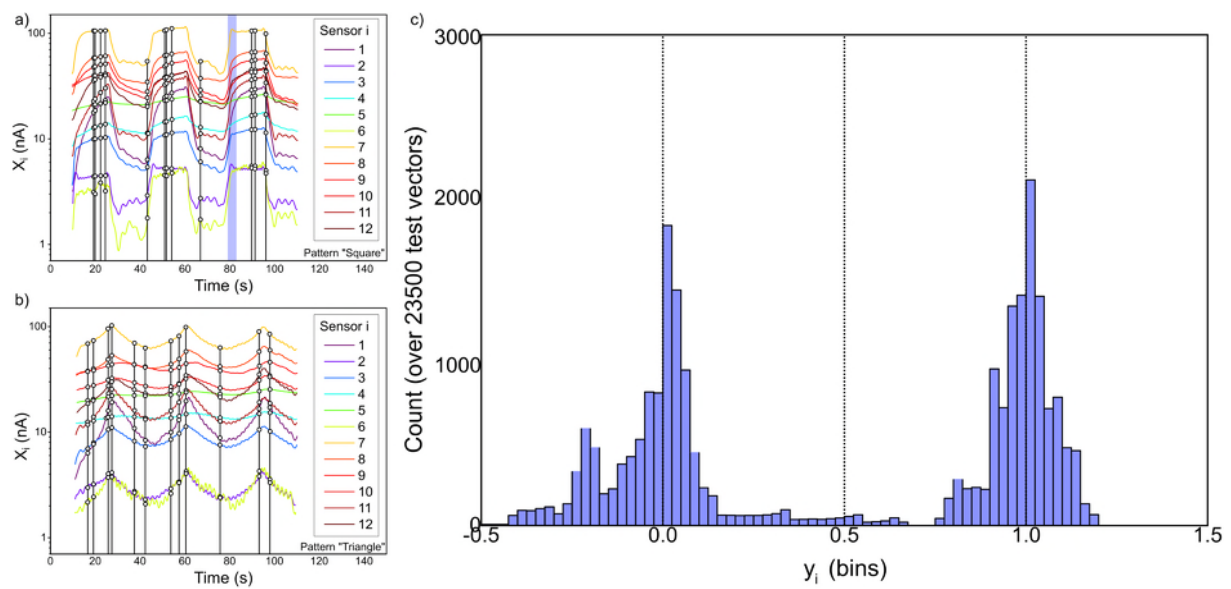

**Figure S2.** *Illustrative example of one iteration for m=12 and n=12 (error rate: 1.12%). 12 training vectors randomly picked from the square data set (a) and the triangle data set (b). The picked up vectors are marked with a vertical lines, overlapped with 12 white open circles. The blue zone in (a) shows the vectors which were mistaken in the test. c, The distribution of the 2x11750 tested vectors, displaying two populations centered on 0 and 1.*

**Table S1.** List of the 10 randomly chosen vectors for each m case, related to the study displayed in Figures 6b and 6c:

| | m 1 | 2 | 3 | 4 | 5 | 6 | 7 | 8 |
|---|---|---|---|---|---|---|---|---|
| 1 | 9 | 4;5 | 11;6;7 | 2;10;3;12 | 9;6;3;2;10 | 3;6;5;11;9;2 | 4;5;3;8;9;7;12 | 11;6;7;2;3;8;1;12 |
| 2 | 7 | 5;3 | 7;5;6 | 7;2;5;12 | 2;4;1;7;5 | 12;5;7;8;10;3 | 5;3;11;2;8;7;9 | 7;5;6;10;11;8;12;2 |
| 3 | 10 | 5;8 | 6;3;10 | 10;8;5;6 | 11;10;1;8;6 | 7;10;1;12;2;4 | 5;8;3;6;2;12;4 | 6;3;10;2;8;9;5;11 |
| 4 | 2 | 5;10 | 6;4;3 | 1;8;3;5 | 8;12;9;10;11 | 1;2;8;4;11;6 | 5;10;8;7;1;11;2 | 6;4;3;9;2;11;8;5 |
| 5 | 7 | 12;2 | 7;11;5 | 10;9;3;1 | 5;8;10;1;4 | 7;5;2;4;8;11 | 12;2;10;1;9;3;11 | 7;11;5;2;9;3;6;8 |
| 6 | 3 | 10;2 | 10;3;8 | 9;8;10;11 | 4;5;9;7;11 | 3;4;5;2;8;9 | 10;2;6;7;1;11;12 | 10;3;8;1;6;11;9;7 |
| 7 | 8 | 11;1 | 10;1;2 | 11;7;10;8 | 11;8;6;7;12 | 8;7;3;9;6;11 | 11;1;8;6;10;7;12 | 10;1;2;5;4;9;12;8 |
| 8 | 7 | 6;2 | 3;2;8 | 6;10;2;7 | 11;3;9;6;1 | 7;10;1;4;5;9 | 6;2;3;12;1;4;9 | 3;2;8;12;1;5;9;10 |
| 9 | 12 | 12;11 | 1;6;9 | 12;2;6;7 | 3;4;12;1;9 | 12;4;8;11;10;3 | 12;11;3;2;10;9;6 | 1;6;9;11;5;3;4;8 |
| 10 | 1 | 2;5 | 7;2;10 | 1;10;3;2 | 3;1;8;12;2 | 1;4;12;8;5;3 | 2;5;1;12;7;8;9 | 7;2;10;12;1;5;8;4 |

| | m 9 | 10 | 11 | 12 |
|---|---|---|---|---|
| 1 | 2;10;3;12;4;7;9;6;5 | 9;6;3;2;10;4;12;8;7;1 | 3;6;5;11;9;2;4;10;12;8;7 | 5;11;9;6;4;3;7;8;2;1;10;12 |
| 2 | 7;2;5;12;3;6;8;11;4 | 2;4;1;7;5;12;3;10;9;6 | 12;5;7;8;10;3;9;4;1;6;2 | 9;3;4;12;8;1;10;6;5;11;7;2 |
| 3 | 10;8;5;6;11;1;3;12;2 | 11;10;1;8;6;4;7;5;9;12 | 7;10;1;12;2;4;6;8;9;11;3 | 4;6;5;1;9;12;3;8;2;10;7;11 |
| 4 | 1;8;3;5;11;10;12;9;4 | 8;12;9;10;11;2;3;1;7;5 | 1;2;8;4;11;6;5;10;7;3;9 | 11;5;10;2;8;1;7;3;4;6;12;9 |
| 5 | 10;9;3;1;11;5;7;8;4 | 5;8;10;1;4;6;3;7;12;9 | 1;10;3;6;8;12;7;4;5;9;2 | 5;1;3;12;6;9;8;2;7;4;11;10 |
| 6 | 9;8;10;11;5;1;12;2;6 | 4;5;9;7;11;1;8;6;12;3 | 3;6;12;10;11;5;4;1;2;7;8 | 1;11;8;9;4;5;12;3;2;6;7;10 |
| 7 | 11;7;10;8;9;2;12;4;1 | 11;8;6;7;12;5;9;1;10;2 | 6;3;8;2;1;10;4;9;7;11;5 | 4;5;8;6;2;3;12;11;1;7;9;10 |
| 8 | 6;10;2;7;11;1;3;12;5 | 11;3;9;6;1;5;2;12;4;8 | 2;6;4;3;7;10;9;12;1;11;5 | 11;5;8;12;9;7;6;10;2;1;4;3 |
| 9 | 12;2;6;7;4;11;10;9;3 | 3;4;12;1;9;10;7;2;11;6 | 8;9;2;10;7;6;4;5;1;3;12 | 10;1;7;3;11;2;9;12;5;8;4;6 |
| 10 | 1;10;3;2;5;9;12;7;11 | 3;1;8;12;2;9;11;6;7;4 | 9;10;4;7;6;1;5;2;3;8;12 | 4;8;11;9;12;6;5;3;10;1;7;2 |

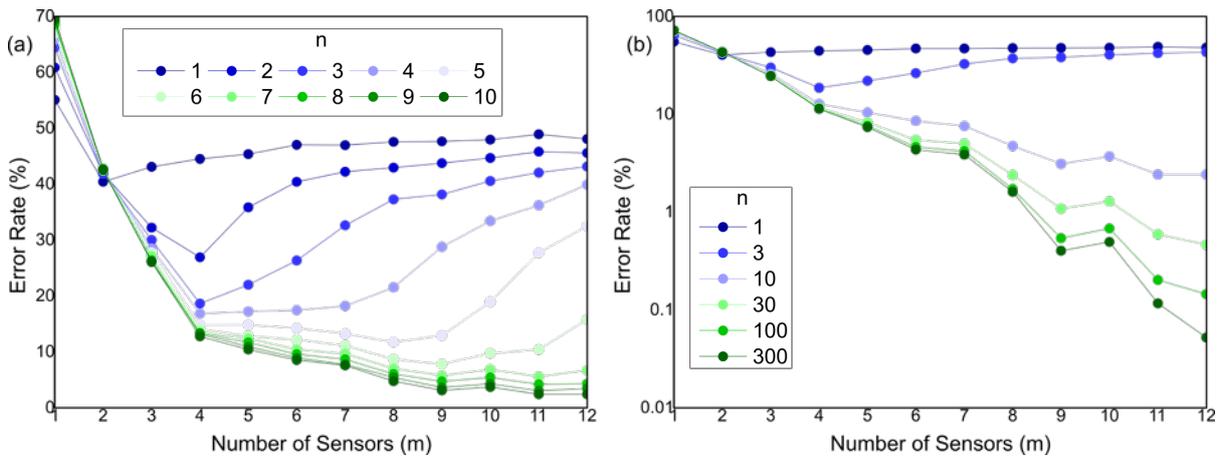

***Figure S3.*** *Averaged performances for different array size m and training vectors n. Average performances as a function of m is obtained by calculating the mean value and deviation on 10 randomly chosen set of vectors for each m value (Table S1). a, linear scale. b, logarithmic scale.*

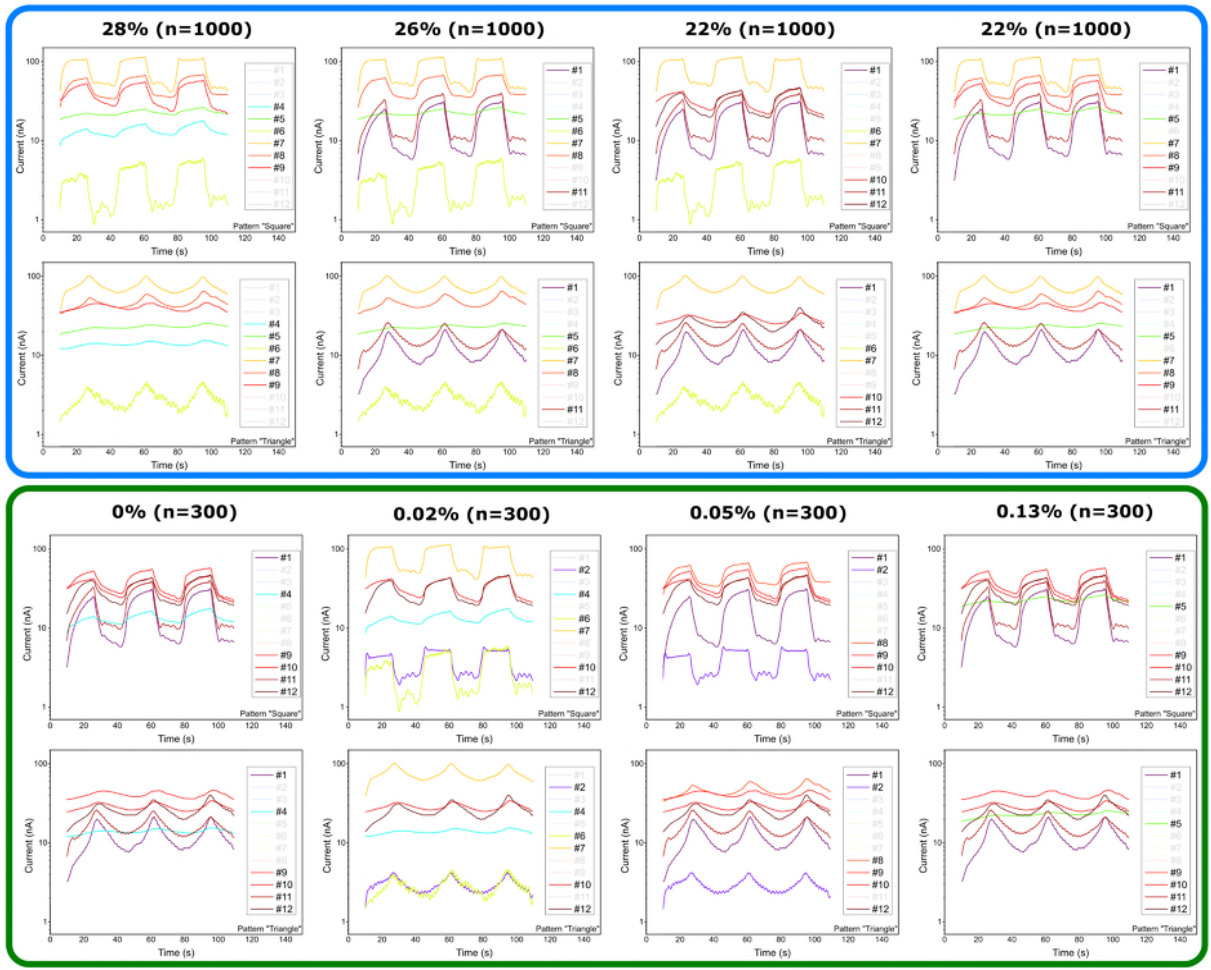

***Figure S4.*** *In the blue frame, the 4 sets of 6 transistors with show the highest error rate in the study plotted in Figure 6a. In the green frame, the 4 sets of 6 transistors with show the lowest error rate in the study plotted in Figure 6a. From this data, it does not appear possible to distinguish what kind of variability promotes the most optimal performances in the transistors network.*

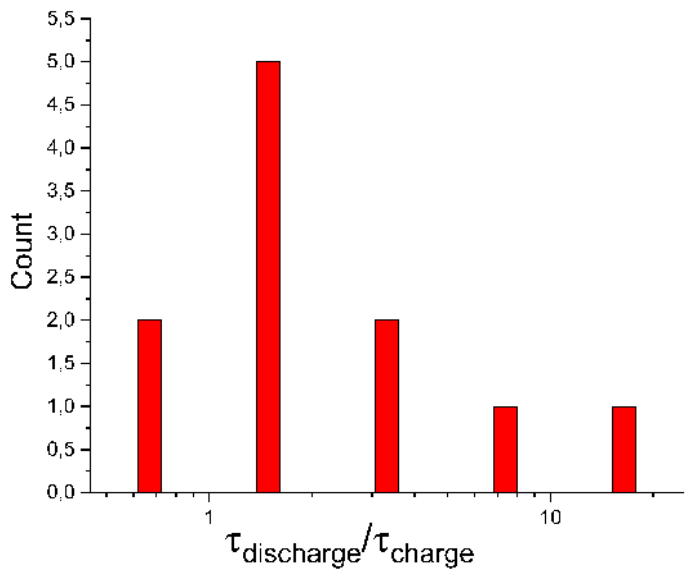

***Figure S5.*** *Statistics for $\tau_{discharge}$ / $\tau_{charge}$. The distribution indicates a clear short-term memory effect (see main text) with $\tau_{discharge}$ / $\tau_{charge}$ <20. Note that the two devices with longer term memory are device #5 and #1 and can lead to both ideal classification or poor classification for n>300 training vectors (see examples in Figure S4). Based on the given distribution of $\tau_{discharge}$/$\tau_{charge}$ ratio, it was not possible to extract some consistent dependency between variability in $\tau_{discharge}$/$\tau_{charge}$ and the classification accuracy.*
*Order of devices from small to high ratio: #9, #4, #7, #8, #12, #6, #3, #10, #11, #5, #1*
*Note that one device (#8) had very short-time constant that cannot be evaluated with the sampling rate used in our protocol.*

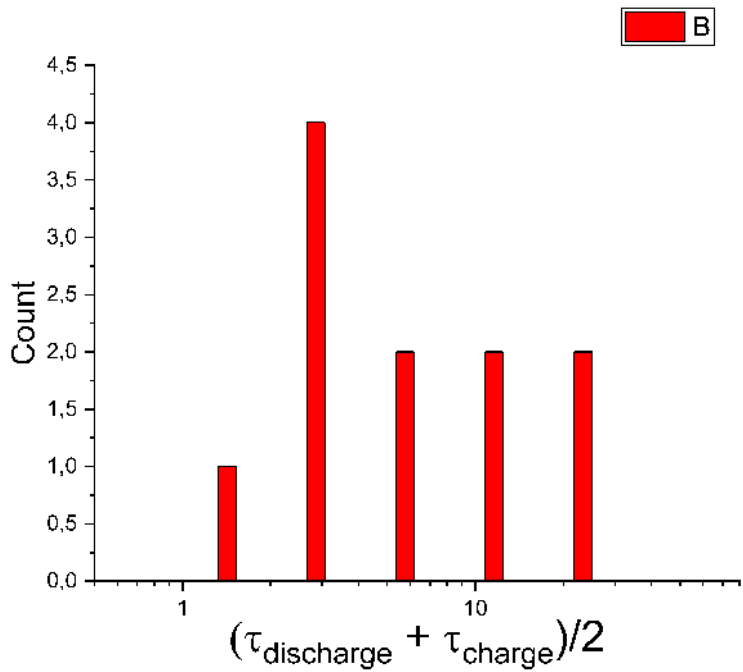

**Figure S6.** *Order of devices from short τ_mean to longer τ_mean: #6, #7, #8, #12, #3, #11, #10, #4, #9, #1, and #5. Note that one device (#8) had very short-time constant that cannot be evaluated with the sampling rate used in our protocol.*

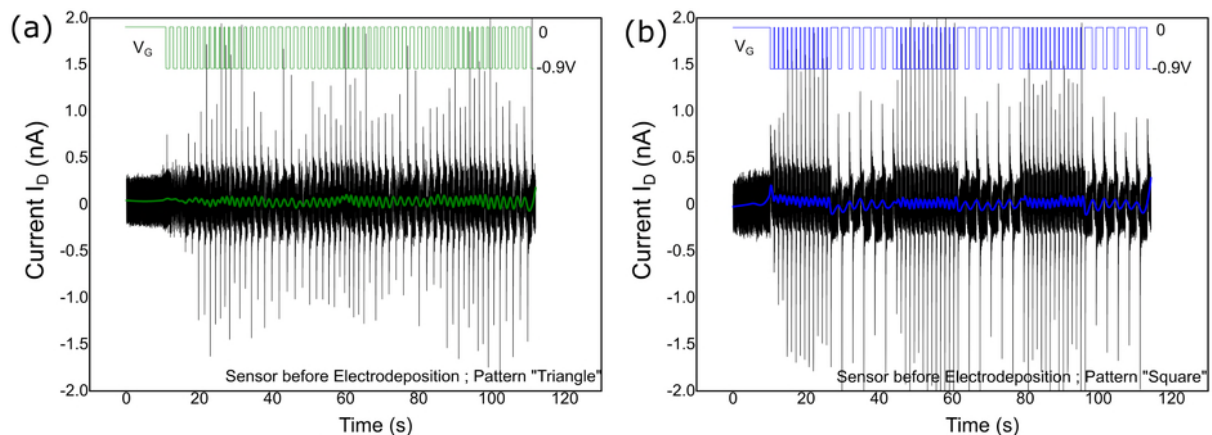

**Figure S7.** *Typical response for the two different patterns (triangle for figure a. and square for figure b.) of an OECT before the electrodeposition of the TEDOT monomer (i.e. the naked source and drain electrodes covered by the electrolyte), showing no signal due to capacitive coupling and testifying of the genuine properties of the device due to the p(TEDOT).*